\documentclass[apj]{emulateapj}
\usepackage{natbib}
\usepackage{ctable}
\slugcomment{}

\shorttitle{On the origin of the dwarf galaxy types}
\shortauthors{Gallart et al.}

\begin{document}

\title{The ACS LCID Project: On the origin of dwarf galaxy types: a manifestation of the halo assembly bias?}

\author{
Carme Gallart\altaffilmark{1,2},
Matteo, Monelli\altaffilmark{1,2},
Lucio Mayer\altaffilmark{3,4},
Antonio Aparicio\altaffilmark{1,2},
Giuseppina Battaglia\altaffilmark{1,2},
Edouard J. Bernard\altaffilmark{5}, 
Santi Cassisi\altaffilmark{6},
Andrew A. Cole\altaffilmark{7},
Andrew E. Dolphin\altaffilmark{8},
Igor Drozdovsky\altaffilmark{1,2},
Sebastian L. Hidalgo\altaffilmark{1,2},
Julio F. Navarro\altaffilmark{9},
Stefania Salvadori\altaffilmark{10,11},
Evan D. Skillman\altaffilmark{12},
Peter B. Stetson\altaffilmark{13},
Daniel R. Weisz\altaffilmark{14,15}}


\altaffiltext{*}{Based on observations made with the NASA/ESA HST,  which is operated by the AURA, under NASA contract NAS5-26555. Observations associated with programs \#8706,\#10505, \#10590}
\altaffiltext{1}{Instituto de Astrof\'{i}sica de Canarias, La Laguna, Tenerife, Spain}
\altaffiltext{2}{Departamento de Astrof\'{i}sica, Universidad de La Laguna, Tenerife, Spain}
\altaffiltext{3}{Institut f\"ur Theoretische Physik, University of Zurich, Z\"urich, Switzerland}
\altaffiltext{4}{Department of Physics, Institut f\"ur Astronomie, ETH Z\"urich,  Switzerland}
\altaffiltext{5}{Institute for Astronomy, University of Edinburgh, UK}
\altaffiltext{6}{INAF-Osservatorio Astronomico di Collurania,Teramo, Italy}
\altaffiltext{7}{School of Physical Sciences, University of Tasmania, Australia}
\altaffiltext{8}{Raytheon, USA}
\altaffiltext{9}{Department of Physics and Astronomy, University of Victoria, Canada}
\altaffiltext{10}{Kapteyn Astronomical Institute, Netherlands}
\altaffiltext{11}{Veni Fellow}
\altaffiltext{12}{Minnesota Institute for Astrophysics, University of Minnesota,  USA}
\altaffiltext{13}{Herzberg Astronomy and Astrophysics, NRC, Canada}
\altaffiltext{14}{Astronomy Department, University of Washington,  USA}
\altaffiltext{15}{Hubble Fellow}






\begin{abstract}

We discuss how knowledge of the whole evolutionary history of dwarf galaxies, 
including details on the early star formation events, can provide insight on the origin 
of the different dwarf galaxy types. We suggest that these types may be imprinted by the early 
conditions of formation rather than being only the result of a recent morphological 
transformation driven by environmental effects.
We  present precise star formation histories of a sample of Local Group 
dwarf galaxies, derived from colour-magnitude
diagrams reaching the oldest main-sequence turnoffs. We argue that these
galaxies can be assigned to two basic types: {\it fast dwarfs\/} that
started their evolution with a dominant and short star formation event, and
{\it slow dwarfs\/} that formed a small fraction of their stars early and have continued 
forming stars until the present time (or almost).
These two different evolutionary paths do not map directly onto the
present-day morphology (dwarf spheroidal vs dwarf irregular). Slow and fast
dwarfs also differ in their inferred past location relative to the Milky
Way and/or M31, which hints that slow dwarfs were generally assembled in
lower density environments than fast dwarfs. We propose that the distinction between a fast and slow dwarf galaxy
reflects primarily the characteristic density of the environment where
they form. At a later stage, interaction with a large host galaxy may play a role in the final gas removal
and ultimate termination of star formation. 


\end{abstract}

\keywords{galaxies: dwarf---galaxies: formation---galaxies: evolution}

\section{Introduction} \label{intro}

The origin of the different dwarf galaxy types and the possible 
evolutionary links between them are the subject of much research and
debate. Dwarf spheroidals (dSph,  devoid of gas and with no star
formation), dwarf irregulars (dIrr, gas rich, star-forming  systems usually
located in the field), and the so-called transition types (dT, with
properties intermediate between the other two) have similarities and
differences  that can yield information on their
possibly linked evolution. On one hand, they obey the same mass-metallicity
relation \citep{Kirby2013}, and follow similar relationships between
central velocity dispersion, core radius, central surface brightness, and
total luminosity \citep{KormendyBender2012}. On the
other hand, they have different gas content and are preferentially found in
different environments, the dSph usually inhabiting denser locations---the
so-called morphology-density relation. This 
classification is based on current properties, which may not reflect past
history, i.e., actual evolution. 

Through such reliable indicators as RR Lyrae variable stars, a bona-fide
old population was routinely found in any dwarf galaxy that was adequately
searched. At early times,
therefore, dwarfs of all types must have been star-forming galaxies. Then,
at some point, {\it some\/} lost their gas and stopped star formation. The
transformation from a star-forming,  dIrr-like galaxy to a dSph galaxy has
been explored by many authors, and the common implicit assumption has
led to the  definition of a ``transition class'' of dwarf galaxies. Even if
there are plausible mechanisms to transform a gas-rich, star-forming dwarf
into a pressure-supported, gas-poor dwarf, a question about the origin of
dwarf galaxy types \citep{SkillmanBender1995} remains:  were the properties
of dwarfs imprinted during their early assembly, or do they result from
events happening later? A crucial piece of information is their very early
star formation history (SFH).

Here, we show that the availability of precise SFHs over the whole lifetime
of a diverse sample of Local Group dwarf galaxies opens the door to an
alternative classification based on  evolution. The {\it
early\/} SFHs of dwarf galaxies can be obtained reliably for the nearest
examples, for which deep color-magnitude diagrams (CMDs) can be obtained
from the ground or using the ACS on  the Hubble Space Telescope (HST).
Observation of the {\it whole\/} main sequence---down to the oldest main
sequence turnoff (oMSTO)---with good photometric accuracy and precision is
essential for obtaining SFHs that include details of the earliest star
formation events \citep{Gallart05ARAA}. The reason is
two-fold. First, on the main sequence---which spans many magnitudes in the
CMD---stars are distributed in a sequence of age as a function of
magnitude that is subject to lower age-metallicity degeneracy than
other CMD regions, where stars of all ages and metallicities occupy
a  narrow interval of color and magnitude. Second, the main sequence is the
best understood phase from theory and, therefore,
SFH determinations are much less affected by model uncertainties.

As high-quality SFHs are being obtained, some cosmological
hydrodynamical simulations on the formation and evolution of dwarf galaxies
are becoming available \citep[][]{Shen2014, Brooks2014, Sawala2015, Onorbe2015arXiv}.   In this context, precise SFHs of dwarf
galaxies with different characteristics and in different environments  can
provide firm observational constraints, sheding new
light into the origin of the different dwarf galaxy types and the
possible relationships between them.

We will analyze the available {\it precise\/} SFHs
for Local Group dwarfs of different types, paying special attention to the
detailed information they provide on their earliest evolution. We will
exclusively discuss galaxies where SFHs have been derived from CMDs
reaching the oMSTOs. This is a fundamental difference compared to other
studies analyzing the SFHs of galaxies with CMDs reaching
different, usually shallower {\it absolute\/} depths \citep{Weisz2014a}. 

\section{New dwarf galaxy types based on full evolutionary histories}

\scriptsize 
\begin{table*}
\begin{center}
\caption {Basic properties of LCID galaxies}\label{properties}
\begin{tabular}{lccccccccc}
\tableline\tableline
\noalign{\vspace{0.1 true cm}}
Galaxy & $M_V$ \tablenotemark{(a)} & r$_h$ \tablenotemark{(a)}&$M_T$\tablenotemark{(g)} & M$_{HI}$\tablenotemark{(a)}& $(m-M)_0$ & R$_{MW}$, V$_{MW} $ \tablenotemark{(a)}& R$_{M31}$, V$_{M31}$ \tablenotemark{(a)}& R$_{LG}$, V$_{LG}$ \tablenotemark{(a)} \\
\noalign{\vspace{0.1 truecm}}
\hline
\noalign{\vspace{0.1 truecm}}
& mag & (') & $(\times 10^6 M_{\odot}$) & $(\times 10^6 M_{\odot}$)& mag &  Kpc, Km s$^{-1}$ & Kpc, Km s$^{-1}$ & Kpc, Km s$^{-1}$ \\
\noalign{\vspace{0.1 truecm}}
\hline
\hline
\noalign{\vspace{0.1 truecm}}
Phoenix  &$-9.9$      &3.76  & $3.2\pm0.3$    & 0.12  &$23.09\pm0.1$ \tablenotemark{(b)}      & 415, -103& 868, -104 &556, -106 \\
LGS3      & $-10.1$   &2.10 &  $1.9\pm0.1$   & 0.38  & $24.07\pm0.15$ \tablenotemark{(c)}    &773, -155 & 269, -43 & 422, -74\\
IC1613   & $-15.2$   &6.81  &  100  \tablenotemark{(a)} & 65     & $24.44\pm0.10$ \tablenotemark{(d)} & 758, -154& 520,  -64 &517,  -90 \\
LeoA      & $-11.7$   &2.15  &   6.0  \tablenotemark{(a)} &  11    &$24.50\pm 0.10$\tablenotemark{(e)}    & 803, -19& 1200, -46 &941, -41\\
Cetus     &$-11.2 $   &3.20  & $7.0\pm0.3$     &  --       & $24.46\pm0.12$ \tablenotemark{(f)}   & 756, -27 & 681,  46 &603,  26 \\
Tucana  & $-9.5   $  &1.10  &$3.2\pm0.1$      &   --      & $24.74\pm0.12$ \tablenotemark{(f)}   &882,  99  & 1355,  62 &1076, 73\\
\noalign{\vspace{0.1 truecm}}
\hline
\hline
\end{tabular}

(a) From \citet{McConnachie2012} and references therein; (b) \citet{Hidalgo09sfhphoenix}; (c) \citet{Hidalgo11sfhlgs3}; (d) \citet{Bernard2010}; (e) \citet{Bernard2013}; (f) \citet{Bernard2009};(g) \citet{Hidalgo2013}

\end{center}
\end{table*}
\normalsize

\subsection{The SFHs of presently isolated dwarfs}\label{isolated}

The LCID\footnote{LCID: Local Cosmology from Isolated Dwarfs project, http://www.iac.es/proyecto/LCID/} project obtained the first CMDs reaching the oMSTOs and precise
SFHs for six relatively isolated Local Group dwarf galaxies (see Table~\ref{properties} for a summary of the properties of these galaxies). The observations 
were designed to achieve a 2 Gyr age resolution at old ages, and this has been achieved as shown by \citet{Hidalgo2013}.  Two dIrr,
IC1613 and Leo~A \citep{Skillman2014, Cole2007};  two dT, LGS3 and Phoenix
\citep{Hidalgo11sfhlgs3, Hidalgo09sfhphoenix}; and two isolated dSph in the
Local Group, Cetus and Tucana \citep{Monelli10sfhcetus,
Monelli10sfhtucana} were studied. The reader is referred to the above papers for details
on the SFH determination.  Subsequently, another dIrr
galaxy, DDO~210, has been studied~\citep{Cole2014}

\begin{figure*} 
\plotone{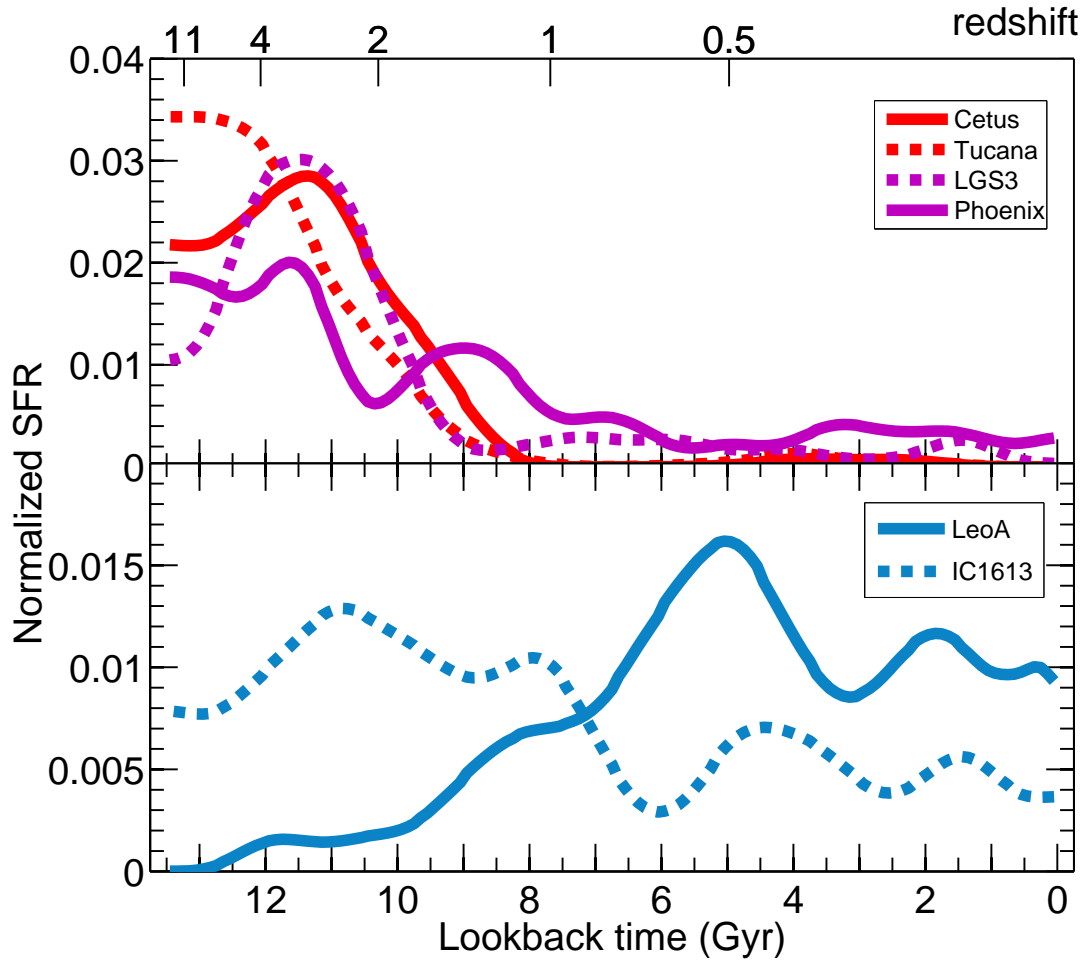}
\caption{Homogeneously derived SFHs for the six LCID galaxies. Upper panel: SFHs for dSph and dT. Lower
panel: SFHs  for the dIrr.} 
\label{sfhs} 
\end{figure*}

Figure~\ref{sfhs} (upper panel) displays the SFHs for the two dSph and the
two dT galaxies of the LCID sample. Tucana and Cetus share the common
characteristic of having formed over 90\% of their stars before 10~Gyr ago,
and host no stars younger than 8--9~Gyr. The SFHs of the two dT galaxies are remarkably similar to
those of the dSphs: they formed over 80\% of their stars
before 9~Gyr ago in spite of having maintained residual 
star formation during the rest of their evolution.  The lower
panel of Figure~\ref{sfhs} displays the SFHs of the two dIrr in our
sample, Leo~A and IC1613. In contrast with the former SFHs, those of the 
dIrrs do not show a dominant early burst of star formation; instead
over 60\% of their stars formed at intermediate and
young ages. 

\subsection{The SFHs of satellite dwarfs} \label{mwdsph}

\begin{figure*} 
\plotone{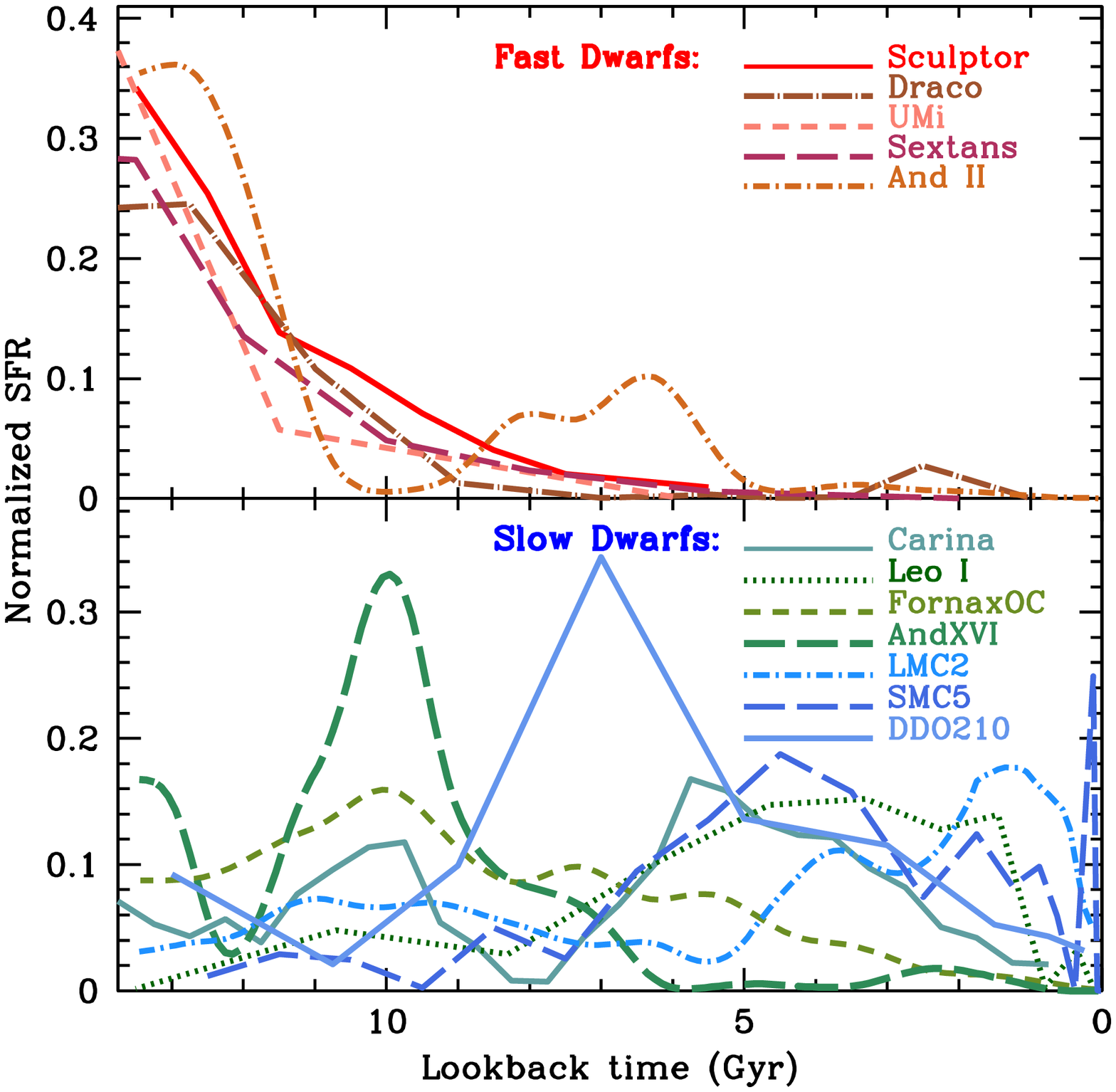}
\caption{SFHs of Local Group fast (upper panel) and slow (lower panel) dwarfs from the literature. Note that SFHs for dwarfs currently classified as dSph (in green shades) and dIrr (in blue shades) are represented in the lower panel.}
\label{literatura} 
\end{figure*}

We now consider dwarfs that are, {\it at present\/}, found close to the
large spirals. Precise SFHs are available for a number of dSph
satellites of the Milky Way \citep[MW,][]{Gallart99sfh, Aparicio01,
Carrera2002, Dolphin2002,  Lee2009, deboer2012scl, delpino2013, deboer2014carina, Weisz2014a}, for the
Magellanic Clouds \citep[e.g.,][]{Noel2009, Cignoni2013, Meschin2014}, and for two M31 satellites: AndXVI and AndII  \citep[][, Monelli et al. in prep]{Weisz2014andies}. We have represented some of these in Figure~\ref{literatura}, together with that of DDO210. Even
though they are not homogeneous among themselves or with the LCID SFHs,
these results generally agree that most MW satellites \citep[UMi, Draco,
Sextans, Scl, CnVI, plus the very faint dwarfs,][]{Brown2014} formed most of their stars before $\simeq$
10~Gyr ago. The more distant dSph satellites, Fornax, Leo~I and Carina show
substantial intermediate-age populations:  their SFHs peaked at ages
younger than 10~Gyr, and most of their star formation occurred at
intermediate ages. In fact {\it the SFHs of these dSph are  similar to
those of dIrr galaxies for most of their lifetimes\/}:
they have low initial star formation rates (SFR) and high SFR at
intermediate ages. The main difference occurs in the last $\simeq$2~Gyr or
less, when they stopped star formation. 
They are classified as dSph for their current properties, but their
history is similar to that of dIrr galaxies.

\subsection{Slow dwarfs and fast dwarfs: a classification based on evolution} \label{newclass}

The availability of precise SFHs reaching the earliest star formation
events for a growing sample of dwarf galaxies enables us to take a 
fresh look into dwarf galaxy types based on evolutionary histories
rather than current properties. Although this is
currently possible only for a limited number of objects in the Local Group,
it provides new insight on possible mechanisms producing
these dwarf galaxy types. 

Based on their full evolutionary histories, we propose that dwarf galaxies
can be grouped in two classes:

$\bullet$ {\it fast dwarfs\/} are those that started
their evolution with a dominant star formation event, but their period of
star formation activity was short ($\lesssim$ few Gyr, see upper panel of Figure~\ref{literatura}); 

$\bullet$ {\it slow dwarfs\/} formed a small fraction of their stellar
mass at an early epoch, and continued forming stars until the present
(or almost, see lower panel of Figure~\ref{literatura}). 

These two evolutionary paths do not map directly onto
the current dwarf. Most notably, some dSphs have
important intermediate-age and young populations, and thus SFHs
that resemble those of dIrr:  in our sample, all dIrr are slow
dwarfs, while some dSph are fast and others are slow.

\begin{figure} 
\includegraphics[scale=0.4]{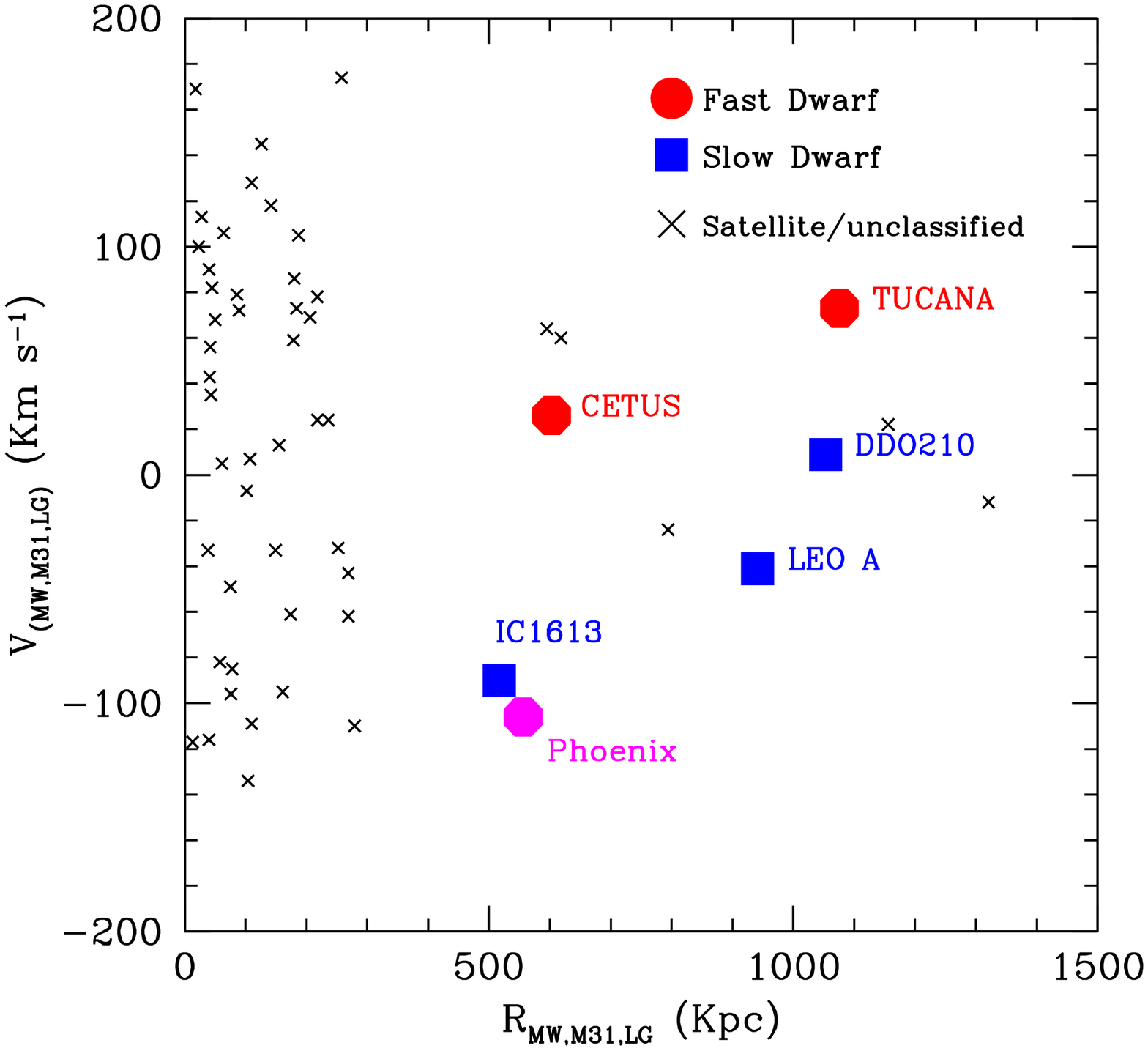}
\caption{Radial velocities  and distances relative to the MW or M31, or to the Local Group barycenter for Local Group dwarf galaxies in the  \citet{McConnachie2012} compilation. For dwarfs located within 300 Kpc of either galaxy (small crosses), the slow and fast dwarfs classification has not been highlighted, since they have an uncertain dynamical history, and  may have orbited more than once about the host. }
\label{global} 
\end{figure}

What else, besides similar SFHs, do slow dwarfs share? The distant
dIrr galaxies with known full SFHs, IC1613, Leo~A and DDO210, are
all located over 400 kpc away from the MW or M31, and have negative (or
small for DDO210) Galactocentric radial velocities,
M31 and the Local Group center of mass (Figure~\ref{global}). Among the closest slow dwarfs, the Magellanic Clouds
and Leo~I have been found to have first entered the MW virial radius
just a few Gyr ago \citep{Sohn2013, Kallivayalil2013}. This might indicate
that all these slow dwarfs {\it assembled\/} in a low density environment and are only recently entering the
Local Group. 

In contrast, most fast dwarfs are close MW satellites, or isolated
dSph (Cetus and Tucana). The latter break the morphology-density relation
in the Local Group. However, they have radial velocities \citep[][see Figure~\ref{global}]{Lewis2007, Fraternali2009} compatible with
their having been close to the Local Group barycenter at early times,
presumably when their assembly was taking place. 

\section{On the origin of the dwarf galaxy types}

Is the role of environment in creating
the different dwarf galaxy types more closely related
to its early influence on the mass assembly process of the dwarf (which is
expected to depend on formation location) than to its effects 
in removing the gas later?

\subsection{Quenching a dwarf: gas removal scenarios}

Most (mostly theoretical) research into the 
transformation from a gas-rich to a gas-poor dwarf galaxy has
focused on determining how dSph galaxies lost
their gas.   \citet{DekelSilk1986, MacLowFerrara1999, Salvadori2008,
Sawala2010}, among others, have explored internal feedback
and the efficiency of gas ejection through supernova-driven outflows. In
general, models indicate that feedback should be able to
totally remove the gas only in extremely low-mass (few $10^6$ M$_\odot$ of
baryonic mass) dwarf galaxies. The inability of feedback alone to
totally remove the gas from currently gas-free galaxies, together with the
existence of a striking morphology-density relation, has led to the current
consensus that environmental processes  such as ram-pressure or tidal stripping by a massive central halo must play
an important role in stripping the gas from dwarf galaxies and halting star
formation \citep[see][for a review]{Mayer2010}.   Finally, including the
effects of an ionizing UV background has been shown to increase the
efficiency of gas loss due to both internal feedback
\citep{SalvadoriFerrara2009,Sawala2010} and ram-pressure stripping \citep{Mayer2010}.

However,  it is important to note that most
MW dSph satellites stopped forming  stars at a very similar (within 2--3~Gyr)
early time ($\simeq$ 10~Gyr ago). If environmental effects were crucial in
removing their gas, one  would expect that they entered the virial
radius of the MW or other massive companion, also at very  {\it similar and early\/} times, for
their SFHs to share the common pattern. Cosmological simulations are
not conclusive on this point. Several studies show that infall times
around $z \sim1$ are most typical inside MW-sized halos
\citep[e.g.,][]{Rocha2012, Wetzel2015}, which would be too late to explain
the early star formation truncation in many fast satellites. However, other
studies of halos that assemble earlier than average, and that lead
to a realistic replica of the MW  \citep{Guedes2011}, find that a fraction of dSphs could be hosted in subhalos
accreting at $z \sim 2-2.5$ \citep{Tomozeiu2015arXiv}.

\subsection{A possible alternative scenario} 

We now examine whether there are theoretical indications that
progenitors of slow and fast dwarfs may be different, and whether this
difference can be linked to the location of the progenitor's halo
when the majority of its mass assembly took place.

As discussed in Section~\ref{newclass}, the current positions and
space motions seem to indicate that slow dwarfs formed
preferentially in lower density environments than fast dwarfs.  This could imply a
delayed assembly history for the dwarf dark matter halos, as
predicted for lower peaks in the field of density fluctuations
\citep[e.g.][]{Lagos2009} and reflected in a number of theoretical investigations that will be discussed next.

\citet{Sawala2014arXiv} present mass-assembly histories for present-day satellites and isolated
dwarfs, in twelve cosmological volumes resembling the Local Group. They
find that, for dark matter sub-halos in the mass range
$M=10^{9-10}M_{\odot}$, which are expected to host the majority of Local
Group dwarfs, the redshift at which a halo progenitor reached 1/2 of its
peak mass was $z_{1/2}\simeq$1.5--2 for field dwarfs and $z_{1/2}\simeq$4
for satellites. \citet{Brooks2014} also find that satellites accrete most of their
mass earlier than field dwarfs and show
an earlier peak in their SFHs than the simulated field dwarfs.
While in our scenario there is no one-to-one correspondence between
satellites and fast dwarfs, or between field galaxies and slow dwarfs, this
correspondence is expected to happen in most cases. Thus,  we expect that the characteristics of fast and slow
dwarfs would emerge in theoretical studies if grouped according 
to their early rather than current locations.

The scenario proposed by \citet{Benitez-Llambay2015}  also fits well within the slow- vs. fast-dwarf hypothesis. They use a high
resolution cosmological simulation of the Local Group to  discuss the dramatic effects of
reionization on the baryonic component  of halos with virial temperatures at
$z_{reion}$ similar to or below $10^4K$. This characteristic 
temperature defines a ``threshold'' mass at $z_{reion}$ that strongly
influences the future SFH of the system:  i) systems collapsing early 
with mass at reionization just above the threshold are able to
form stars before reionization, but their star formation is abruptly
truncated by the combined effects of reionization and feedback from the
early stellar population. They are usually characterized by a population of
old stars. ii) In halos with masses below the threshold for star
formation at $z_{reion}$ and gas densities high enough to survive
photoevaporation due to self-shielding, star formation can instead be 
delayed  and only (re-)start at
later times \citep[e.g.][]{Ricotti2009}, e.g. when the host halo
becomes massive enough to allow some of the gas to
cool and fragment.  These two types of systems can be
associated with fast and slow dwarfs respectively. Finally, in their
simulation of a sample of seven dwarf galaxies in a low density
environment, \citet{Shen2014} find that, although above a virial mass of
$10^9 M_{\odot}$ star formation commences quite early ($z > 4$), all of
their dwarfs would be classified as slow. 

These scenarios for the formation and evolution of slow and fast dwarfs are
still compatible with final star formation shutdown when a slow dwarf enters the host halo area of influence. In fact,
tidal effects could play a role in the final removal of gas at
late times in some slow MW satellites, such as Leo~I or Fornax,
which stopped their star formation only recently, as opposed to field
isolated slow dwarfs, like Leo~A or IC1613, that are still forming stars and
retain sizable amounts of gas. 

\section{Summary, conclusions and outlook}

We have discussed the properties of the subsample of
Local Group dwarfs with precise life-time SFHs, derived from CMDs
reaching the oMSTOs. These SFHs reveal two distinct
evolutionary paths that do not lead directly to the current morphological
classification (in dSph, dIrr and dT). One evolutionary path is
characterized by an initial dominant and short (few Gyr) star forming
event, with little or no star formation thereafter. The other leads to dwarf
galaxies dominated by intermediate-age populations, that have continued
forming stars until the present (or almost). We have called the galaxies
displaying these evolutionary paths {\it fast\/}  and {\it slow\/} dwarfs,
respectively. In our sample, all dIrr are slow dwarfs, while some
dSph are fast and others are slow. In addition to SFHs, slow and fast
dwarfs also differ in their inferred early location relative to the local
large galaxies: as opposed to fast dwarfs, slow dwarfs' positions and
radial velocities are compatible with a late first infall into the Local Group,
which would imply that they were assembled in lower
density environments than fast dwarfs.

We thus suggest that the nature of fast or
slow dwarfs is determined early, depending on the formation
conditions of the galaxy. The progenitor halos of fast dwarfs 
assembled early and quickly in high density environments, where 
interactions triggering star formation were common, likely leading to
high star-formation rates even before reionization. Strong gas loss would
follow as a consequence of the effects of reionization and feedback acting
together. Slow dwarfs resulted from delayed, slower mass assembly
occurring in lower density environments, which in turn 
led to a delayed onset of star formation, occurring when the halo had
grown massive enough to allow the gas to cool and form stars. This 
implies milder feedback and gas loss, and the possibility to keep forming
stars on a long timescale. A strong interaction with a large host could
play a role in the late, final removal of gas from the dwarf
galaxies that infall late. The morphology-density
relation observed (with exceptions) in Local Group dwarf galaxies
today would thus be a consequence of their 
{\bf formation}
in more or less dense environments around the Local Group.  

It is interesting to note that the proposed scenario can be seen as a
consequence of the so-called halo assembly bias \citep{Gao2005}: at the
same halo mass, halos that assemble earlier cluster more than halos that
assemble later, hence automatically evolve in a higher density environment.
Then if baryons---as expected---trace the assembly of the dark halo, it is
conceivable that more clustered dwarfs will also be faster at assembling
their stellar component. This is because gas feeding will be more efficient
in a denser region of the cosmic web, presumably achieving earlier the 
density required for efficient star formation. The effect of
assembly bias might be amplified at the scale of dwarfs because
their shallow potential well makes them  more sensitive to processes that
can keep gas density low enough to prevent  radiative cooling and star
formation, such as the cosmic ionizing background. Likewise, star
formation of dwarfs born near massive  halos
could be terminated earlier than in similar dwarfs  
formed in "average cosmic regions", because of the effect of {\it
local\/} radiative feedback.

In our scenario, we expect that the very faint galaxies
that may be discovered far from the MW or M31 in forthcoming deep
photometric surveys, such as with the LSST, will not 
necessarily be extremely old  galaxies like the faintest dwarfs close
to the MW, but some may contain relatively young stellar populations. 
They may have started star formation late at a low rate, and being less 
affected by reionization and internal feedback, may have undergone a very
extended SFH despite their very low mass. A few examples of such
galaxies have already been found, like Leo~P
\citep{McQuinn2013} or Leo~T \citep{Weisz2012}.

To gain further insight regarding the hypothesis presented in this paper it is extremely 
important to increase the sample of distant Local Group dwarf galaxies with oMSTO 
photometry available over a significant fraction of their body; in the near future this will 
only be possible with the ACS on HST. Additionally, knowledge of the orbits of Local 
Group galaxies will allow us to contrast our hypothesis by better constraining 
the type of environment in which they were formed. Finally, this scenario may be explored with 
current and forthcoming cosmological hydrodynamic simulations by taking a slightly different 
vantage point, that is, by grouping samples of simulated dwarfs according to their early location 
when assembling their mass, rather than according to their present location as satellites or 
field dwarfs.

\normalsize

\acknowledgments    C.G., M.M., A.A. and S.H. acknowledge support from grants  AYA2013-42781 and AYA2014-56795 S.S. thanks NWO for her VENI grant 639.041.233. DRW is supported by NASA Hubble Fellowship grant HST-HF-51331.01. G.B is supported by the Spanish MINECO under RYC-2012-11537. SC acknowledges financial support by PRIN-MIUR (2010LY5N2T). 


\end{document}